\documentclass[pdftex,twocolumn,10pt,letterpaper]{extarticle}
\usepackage{balance}
\usepackage{nicefrac}
\usepackage{url}

\newcommand{\AUTHORS}{Sam Burnett, Nick Feamster}
\newcommand{\TITLE}{\name{}: Lightweight Measurement of \\ Web Censorship
 with Cross-Origin Requests}
\newcommand{\KEYWORDS}{Put your keywords here}
\newcommand{\CONFERENCE}{Somewhere}
\newcommand{\PAGENUMBERS}{yes}       % "yes" or "no"

\newcommand{\showComments}{yes}
\newcommand{\comment}[1]{}
\newcommand{\onlyAbstract}{no}

\usepackage{ifthen}
\ifthenelse{\equal{\PAGENUMBERS}{yes}}{%
\usepackage[nohead,
            left=1.0in,right=1.0in,top=1.0in,
            footskip=0.5in,bottom=1.0in,     % Room for page numbers
            columnsep=0.25in
            ]{geometry}
}{%
\usepackage[noheadfoot,left=1.0in,right=1.0in,top=1.0in,
            footskip=0.5in,bottom=1.0in,
            columnsep=0.25in
	    ]{geometry}
}

\usepackage[font=small,format=plain,labelfont=bf,textfont=it,up]{caption}

\usepackage[compact]{titlesec}              % SPACE
\usepackage{paralist}
\usepackage{enumitem}
\setlist{itemsep=0pt,parsep=0pt}             % more compact lists

\renewcommand{\paragraph}[1]{\vspace*{0.03in}\noindent{\bf #1}}

\newcommand{\tr}[1]{}

\usepackage{fancyhdr}

\ifthenelse{\equal{\PAGENUMBERS}{yes}}{%
  \pagestyle{plain}
}{%
  \pagestyle{empty}
}

\usepackage{cite}

\usepackage{booktabs}
\usepackage{color}
\usepackage{colortbl}
\usepackage{float}                           % Must appear before hyperref to
\usepackage{mathptmx}                        % Times/Times-like math symbols
\usepackage{courier}
\usepackage[scaled=0.92]{helvet}

\usepackage[breaklinks=true,pagebackref,filecolor=black,citecolor=blue,urlcolor=black,linkcolor=blue,colorlinks,pdfpagelabels,pdfpagelayout=SinglePage]{hyperref}

\renewcommand*{\backref}[1]{}
\renewcommand*{\backrefalt}[4]{
  \ifcase #1 %
    (Not cited.) %
  \or
    {\small\sffamily (Cited on page~#2.)}%
  \else
    {\small\sffamily (Cited on pages~#2.)}%
  \fi
}

\hypersetup{%
pdfauthor = {\AUTHORS},
pdfsubject = {\CONFERENCE},
pdfkeywords = {\KEYWORDS},
bookmarksopen = {true}
}

\definecolor{placeholderbg}{rgb}{0.85,0.85,0.85}

\usepackage[pdftex]{graphicx}
\usepackage{soul}
\usepackage{subcaption}
\usepackage{amsthm}
\usepackage{amssymb}
\usepackage{listings}
\usepackage{tikz}
\usetikzlibrary{arrows}

\usepackage[absolute]{textpos}

\newfloat{acmcr}{b}{acmcr}

\newcommand{\note}[2]{
    \ifthenelse{\equal{\showComments}{yes}}{\textcolor{#1}{#2}}{}
}

\newcommand{\ie}{{\em i.e.}}
\newcommand{\eg}{{\em e.g.}}

\newcommand{\fp}{\vspace*{0.05in}\noindent}
\newcommand{\name}{Encore}

\newtheoremstyle{tighter}
    {3pt}        % Space above
    {3pt}        % Space below
    {\em}        % Body font
    {}           % Indentation
    {\bfseries}  % Head font
    {:}          % Head punctuation
    {.5em}       % Head space
    {}           % Custom head spec
\theoremstyle{tighter}

\date{}
\title{\textbf{\TITLE}}
\author{
Sam Burnett \\
{\em Georgia Tech} \\
{\sf sam.burnett@gatech.edu}
\and
Nick Feamster \\
{\em Princeton} \\
{\sf feamster@cs.princeton.edu}
}

\usepackage{microtype}  % SPACE

\begin{document}

\maketitle

%\AcmCopyright
%\ToAppear

\begin{sloppypar}
\begin{abstract}
  Despite the pervasiveness of Internet censorship, we have scant data
  on its extent, mechanisms, and evolution.  Measuring censorship is
  challenging: it requires continual measurement of reachability to many
  target sites from diverse vantage points. Amassing suitable
  vantage points for longitudinal measurement is difficult; existing
  systems have achieved only small, short-lived deployments.  We
  observe, however, that most Internet users access content via Web
  browsers, and the very nature of Web site design allows browsers to
  make requests to domains with different origins than the main Web
  page. We present \name{}, a system that harnesses cross-origin
  requests to measure Web filtering from a diverse set of vantage points
  without requiring users to install custom software, enabling
  longitudinal measurements from many vantage points. We
  explain how \name{} induces Web clients to perform cross-origin
  requests that measure Web filtering, design a distributed platform
  for scheduling and collecting these measurements, show the
  feasibility of a global-scale deployment with a pilot study and an
  analysis of potentially censored Web content, identify several cases
  of filtering in six months of measurements, and discuss ethical
  concerns that would arise with widespread deployment.
\end{abstract}

\ifthenelse{\equal{\onlyAbstract}{no}}{%
\section{Introduction}\label{sec:intro}

Internet censorship is pervasive: by some estimates, nearly 60 countries
restrict Internet communication in some way~\cite{www-oni}. As more citizens in
countries with historically repressive governments gain Internet access,
government controls are likely to increase. Collecting pervasive, longitudinal
measurements that capture the evolving nature and extent of Internet censorship
is more important than ever.

Researchers, activists, and citizens aim to understand what, where, when, and
how governments and organizations implement Internet censorship. This knowledge
can shed light on government censorship policies and guide the development of
new circumvention techniques.  Although drastic actions such as introducing
country-wide outages (as has occurred in Libya, Syria, and Egypt) are eminently
observable, the most common forms of Internet censorship are more subtle and
challenging to measure. Censorship typically targets specific
domains, URLs, keywords, or content; varies over time in response to
changing social or political conditions (\eg, a national election); and can be
indistinguishable from application errors or poor performance (\eg, high
delay or packet loss). Detecting more nuanced forms of censorship requires
frequent measurement from many varied vantage points.

Unfortunately, consistently and reliably gathering these types of measurements
is extremely difficult. Perhaps the biggest obstacle entails obtaining access to
a diverse, globally distributed set of vantage points, particularly in the
regions most likely to experience censorship. Achieving widespread deployment in
these locations often requires surmounting language and cultural barriers and
convincing users to install measurement software. Although researchers have
begun to develop custom tools to measure filtering (\eg,
OONI~\cite{Filasto:foci2012,www-ooni}, Centinel~\cite{www-centinel}), widespread
deployment remains a challenge. Instead, researchers have resorted to
informal data collection (\eg, user reports~\cite{www-herdict}) or collection
from a small number of non-representative vantage points (\eg, PlanetLab
nodes~\cite{sfakianakis:2011}, hosts on virtual private networks, or even
one-off deployments of single vantage points) that might not observe the same
filtering that typical users experience.

This paper takes an alternate approach: rather than ask each user to deploy
custom censorship measurement software, we use existing features of the Web to
induce unmodified browsers to measure Web censorship. Many users
access the Internet with a Web browser, so inducing these browsers to perform
censorship measurements will enable us to collect data from a larger, more
diverse, and more representative set of vantage points than is possible with
custom censorship measurement tools.

Our system, \name{}, uses Web browsers on nearly every
Internet-connected device as potential vantage points for collecting
data about what, where, and when Web filtering occurs. \name{} relies on
a relatively small number of Web site operators ({\em webmasters}) to
install a one-line embedded script that attempts to retrieve content
from third-party Web sites using cross-origin
requests~\cite{www-mozilla-cor,www-google-cor}. The \name{} script
induces every visitor of these modified pages to request an object from
a URL that \name{} wishes to test for filtering.  Although same-origin
policies in browsers prohibit many kinds of requests (\eg, to thwart
cross-site request forgery), we demonstrate that the cross-origin
requests that browsers {\em do} allow are sufficient to collect
information and draw conclusions about Web filtering.  A major
contribution of our work is to show that meaningful conclusions
about Web filtering can be drawn from the side channels that exist in
cross-origin requests.
%\name{} then submits feedback about
%whether that request succeeded for correlation across vantage points.

\name{}'s simplicity comes at the cost of significant limitations on the
types of measurements it can collect and the conclusions we can draw
from its measurements. First, \name{}'s measurements must operate within
the constraints of the cross-origin requests that Web browsers
permit. For example, the {\tt img} HTML directive yields the most
conclusive feedback about whether an object fails to load, but it can
only be used to test images, not general URLs. This limitation means
that while it may be useful for detecting (say) the filtering of an
entire DNS domain, it cannot test the reachability of specific
(non-image) URLs. \name{}'s design must recognize which cross-origin
requests browsers permit and use combinations of these requests to draw
inferences with higher confidence. Second, because \name{} requires
webmasters to augment their existing Web pages, \name{} must be easy to
install and incur minimal performance overhead on the Web sites where it
is deployed. Finally, great care is required when measuring censorship
because accessing sensitive sites may endanger users in repressive
countries. Our research focuses on \name{}'s design and implementation,
and is not a measurement study {\em per se}. 

\name{} can detect {\em whether} certain URLs are filtered, but it cannot
determine {\em how} they are filtered. Subtle forms of filtering (\eg,
degrading performance by introducing latency or packet loss) are difficult to detect,
and detecting content manipulation (\eg, replacing a Web page with
a block page, or substituting content) using \name{} is nearly impossible.  
Thus, \name{} may complement other censorship measurement systems, which can
perform detailed analysis but face much higher deployment hurdles. Ultimately,
neither \name{} nor other censorship analysis tools can determine human
motivations behind filtering, or even whether filtering was intentional; they
only provide data to policy experts who make such judgments.

\section{Related Work}\label{sec:related}

We summarize existing censorship measurement techniques and previous
studies of Internet censorship; other policy reports
of Internet censorship (which can ultimately seed our measurements); and
other efforts to perform measurements from clients using advertisements
or embedded images. Although we broadly survey Internet censorship
practices, \name{} focuses on Web filtering.

\paragraph{Censorship measurement tools.}  The prevailing mode for measuring
Internet censorship is to develop custom measurement software and identify users
who are willing to either install the software or otherwise host a measurement
device that runs the software.  Existing measurement tools include
OONI~\cite{Filasto:foci2012}, Centinel~\cite{www-centinel}, and
CensMon~\cite{sfakianakis:2011}. Both OONI and Centinel can be deployed on
end hosts.  These tools perform much more detailed analysis of {\em how} censors
implement blocking, but to date both have seen only limited deployment, likely
because they require recruitment of users who are willing to install and
maintain the measurement software.  CensMon was only deployed for a brief period
on PlanetLab, a global network of servers hosted in academic networks; such
measurements are unlikely to be representative, as residential and mobile
broadband networks can face much different censorship practices than academic
and research networks~\cite{nabi2013anatomy,xu2011pam}.  At this point, we are
not aware of any censorship measurement system that continuously collects
measurements from a global set of vantage points; this is the gap that \name{}
aims to fill.

\paragraph{Censorship measurement studies.}  Several researchers have performed
``first look'' studies of censorship in various countries such as
Pakistan~\cite{nabi2013anatomy}, Iran~\cite{aryan2013internet}, and
China~\cite{zittrain2003internet,Crandall2007,Clayton2006}.  Zittrain {\em et
al.}'s study of censorship in China~\cite{zittrain2003internet} performed Web
requests to hundreds of thousands of sites, but did so from only a handful of
dialup modems that the authors deployed.
%It is now outdated and deserves fresh
%measurements (for example, the report claims that the most frequent form of
%censorship in China is IP-based blocking, and that keyword-based filtering is
%rarely used).
Crandall~\cite{Crandall2007}, Clayton~\cite{Clayton2006},
and Ensafi~\cite{ensafi2014detecting}
exploit symmetric behavior of the Chinese firewall to measure it from clients
outside China; such measurements are easier to collect than \name{}, but
the technique does not work in all countries.
%Ultimately, the measurements were aborted, as all of the authors' modems in
%China were prevented from connecting.
The studies of censorship in Iran and Pakistan were more limited: the Iran study
apparently performed measurements from a single vantage point for only two
months~\cite{aryan2013internet}, and the Pakistan study performed measurements
from only five test networks over about two months~\cite{nabi2013anatomy}. Each
of these studies offers a useful snapshot into a country's filtering practices
at a particular point in time, but data collection is neither widespread nor
continuous. The OpenNet Initiative has conducted the only long-term study to
date, but its data collection is sporadic, making it difficult to compare
filtering practices across countries and time~\cite{www-oni}.
%Although these studies often provide more detailed analysis about censorship
%practices within a single country, they do not run continuously, and they do
%not offer global coverage; indeed, even within a single country, coverage is
%fairly limited and transient.

\paragraph{Sources of block lists.} Some policy organizations publish
reports concerning censorship practices around the world.  For example,
the Open Network Initiative routinely publishes qualitative reports
based on measurements from a limited number of vantage points, with
scant insight into how censorship evolves over short timescales or what
exactly is being filtered~\cite{www-oni-research,www-oni-china}.  Other
projects such as Herdict~\cite{www-herdict},
GreatFire~\cite{www-greatfire}, and Filbaan~\cite{www-filbaan} maintain
lists of domains that may be blocked.  Herdict compiles reports from
users about domains that are blocked from a certain location; such
reports lack independent verification.  GreatFire monitors reachability
of domains and services from a site behind China's censorship firewall;
it also maintains historical measurement results.  Each tool offers
limited information driven by user-initiated reporting or measurements,
yet these services and reports can serve as initial lists of URLs to
test using \name{}.

\paragraph{Cross-origin requests for client measurement.}  Bortz {\em et al.}
use timing information from cross-site requests to infer various information,
such as whether a user is logged into a particular site or whether a user has
previously visited a Web page~\cite{bortz2007exposing}. Karir {\em et al.} use
embedded Javascript with cross-origin requests to measure IPv6 reachability and
performance from large numbers of clients~\cite{karir2013understanding}; the use
of embedded cross-origin requests to obtain large number of clients is similar
to \name{}'s design.  Other systems have used cross-origin requests to third
parties to determine information such as network latency between a client and
some other Internet destination~\cite{gummadi2002king,www-noction}. In
particular, Casado and Freedman quantified the prevalence of clients behind NATs
and proxies by delivering measurement code to clients in a manner very similar
to \name{}~\cite{casado2007peering}. Puppetnets exploits weaknesses in browser
security to coerce browsers to unwittingly participate in denial-of-service
attacks~\cite{lam2006puppetnets}.
These tools use similar techniques as \name{}, but they primarily
aim to measure network performance or past user behavior based on the timing of
successful cross-origin requests.  They do not infer
reachability of domains, IP addresses, or URLs based on the success (or lack
thereof) of cross-origin requests.

% Concept doppler
% Previous Zittrain studies

% - Related work
%   - Censorship measurement techniques and studies
%     - "Empirical study of a national-scale distributed intrusion detection system: Backbone-level filtering of html responses in china." ICDCS 2010.
%     - "CensMon: A Web censorship monitor." FOCI 2011.
%     - OONI
%     - "The Anatomy of Web Censorship in Pakistan." FOCI 2013
%     - "Internet Censorship in Iran: A First Look." FOCI 2013
%     - "Internet filtering in China." IEEE Internet Computing 2003
%   - Complementary work: where to get block lists
%     - GreatFire, Filbaan, Herdict, ONI reports, etc.
%   - Tools which could measure or frustrate our measurements
%     - Ghostery, AdBlock, NoScript, RequestPolicy
%   - Using ads to measure from many clients:
%     - "Measuring IPv6 with advertisements for fun and profit." AIMS 2012.
%     - "Robust Defenses for Cross-Site Request Forgery." CCS 2008.
%     - "Protecting Browsers from DNS Rebinding Attacks." CCS 2007.
%   - Other work on cross-origin requests
%     - "Exposing Private Information by Timing Web Applications." WWW 2007.
%     - Some stuff about Web bugs
%     - Privacy stuff from Bala?
%     - Analytics
%       - NetVMG, Sockeye, RouteScience, Google, etc.
%     - "King: Estimating Latency between Arbitrary Internet End Hosts." IMW 2002.

\section{Background}\label{sec:background}

We discuss Web filtering and threats that may interfere with attempts to measure
it. We also explain cross-origin requests.

\subsection{Threat Model}

To implement Web filtering, smaller countries often have centralized traffic
filters on a national backbone; larger countries require each ISP to implement a
censorship policy; some countries, like China, do both~\cite{xu2011pam}. Web
filtering typically takes place when the client performs an initial DNS lookup
(at which point the DNS request may result in blocking or redirection), when the
client attempts to establish a TCP connection to the Web server hosting the
content (at which point packets may be dropped or the connection may be reset),
or in response to a specific HTTP request or response (at which point the censor
may reset the TCP connection, drop HTTP requests, or redirect the client to a
block page).

\tr{
\name{} can usually determine when a page is blocked,
but in most cases cannot determine how. Section~\ref{sec:inference} details
\name{}'s inference algorithm for detecting various forms of Web filtering using
cross-origin requests.
}

Our goal is to observe instances of Web filtering and report them to a central
authority (\eg, researchers) for analysis. We assume an adversary that can
reject, block, or modify any stage of a Web connection in order to filter Web
access for subsets of clients, although we assume the adversary uses a blacklist
and is unwilling to filter all Web traffic, or even significant fractions of all
Web traffic.  This adversary influences \name{}'s design in three ways: (1) the
main goal of \name{} is to measure this adversary's Web filtering behavior; (2)
the adversary may attempt to filter clients' access to \name{} itself, thereby
preventing them from collecting or contributing measurements; and (3) the
adversary may attempt to distort \name{}'s filtering measurements by allowing
measurement traffic but denying other access to the same site. This paper
considers all three aspects of the adversary. \tr{We also consider that an
adversary might block, corrupt, or falsify either the measurement tasks we
distribute to clients or the measurement results we collect from them.}

\subsection{Cross-Origin Requests}

% - Primer on cross-origin request vulnerabilities
%   - Web pages can reference other files, for example, images, scripts, style
%     sheets, etc.
%   - Every Web page has an origin, which is typically defined as ...
%   - Different kinds of resources have restrictions on how sites may load them.
%   - Some kinds of objects can be loaded or referenced across domains without restriction.
%   - Browsers restrict some kinds of references for security
%     - For example, restrict access to scripts to prevent cross-site request forgery
%   - Even with restrictions, leaks are possible.
%   - Different browsers might have different policies / vulnerabilities
%   - We use both explicitly allowed cross-origin requests and
%     information leaks in restricted request types to measure
%     accessibility of third party Web resources.

Web browsers' {\em same-origin policies} restrict how a Web page from one origin
can interact with resources from another; an {\em origin} is defined as the
protocol, port, and DNS domain (``host'')~\cite{www-mozilla-cor}. In general,
sites can send information to another origin using links, redirects, and form
submissions, but they cannot receive data from another origin; in particular,
browsers restrict cross-origin reads from scripts to prevent attacks such as
cross-site request forgery.  However, cross-origin {\em embedding} is typically
allowed and can leak some read access. The cornerstone of \name{}'s design is to
use information leaked by cross-origin embedding to determine whether a
client can successfully load objects from another origin.

Various mechanisms allow Web pages to embed remote resources using HTTP
requests across origins; some forms of cross-origin embedding are not
subject to the same types of security checks as other cross-origin
reads.  Examples of resources that can be embedded include simple markup
of images or other media (\eg, {\tt <img>}), remote scripts (\eg, {\tt
  <script>}), remote stylesheets (\eg, {\tt <link rel="stylesheet"
  href="...">}), embedded objects and applets (\eg, {\tt <embed
  src="...">}), and document embedded frames such as iframes (\eg, {\tt
  <iframe>}).  Each of these remote resources has different restrictions
on how the origin page can load them and hence leak different levels of
information.  For example, images embedded with the {\tt img} tag
trigger an {\tt onload} event if the browser successfully retrieves and
renders the image, and an {\tt onerror} event otherwise.  The ability
for the origin page to see these events allows the origin
page to infer whether the cross-origin request succeeded.
%, even though
%the request refers to an object from a different origin.

Although cross-origin embedding of media provides the most explicit feedback to
the origin about whether the page load succeeded, other embedded references can
still provide more limited information, through timing of {\tt onload}
invocation or introspection on a Web page's style.  Additionally, browsers have
different security policies and vulnerabilities; for example, we discovered that
the Chrome browser allows an origin site to load any cross-origin object via the
{\tt script} tag, which allows us to conduct a much more liberal set of
measurements from Chrome.  One challenge in designing \name{} is
determining whether (and how) various embedded object references can help infer
information about whether an object was retrieved successfully.

\section{\mbox{Measuring Filtering with \name{}}}\label{sec:inference}

This section explains how \name{} measures Web filtering using
cross-origin requests. 
\tr{
We first provide an overview of how \name{}
works, and describe the specific limitations on what \name{} can
and cannot conclude from the information it collects.
We then describe the measurement tasks that \name{} can
perform, and what it can (and cannot) infer from these measurements.
Finally, we describe the setup that we use to validate \name{}'s
measurements.  Section~\ref{sec:system} builds a distributed Web
filtering measurement platform using the measurement and inference
primitives that we develop in this section. We defer discussion of how
origin Web servers learn which measurement targets to test from each
client ({\em target selection}) and how Web clients send the results of
their measurements to a central location for analysis ({\em measurement
  collection}) to Section~\ref{sec:system}.
}

% seemed repetitive, so I cut this. - NF
%We explore how to detect whether Web clients in a particular country,
%organization, or ISP can access a specific Web resource or set of Web
%resources. Web resources are anything identified by an HTTP URL, and are
%typically Web pages, image files, style sheets, or flash applets.

\begin{figure}
  \includegraphics{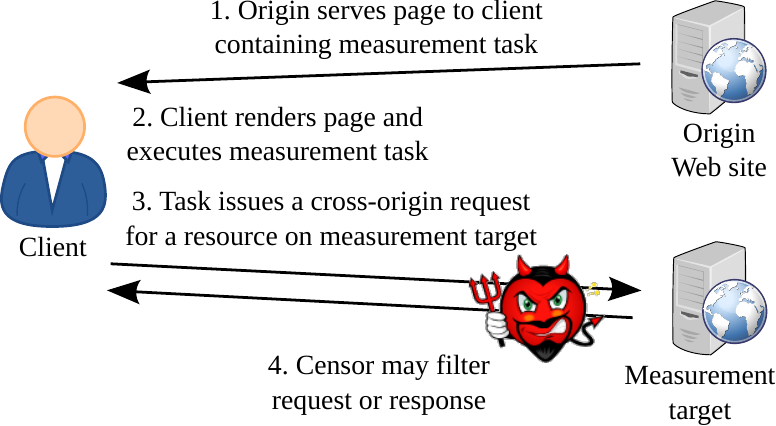}
  \caption{\name{} induces browsers to collect Web filtering measurements by
  bundling measurement tasks inside pages served by an origin Web site.}
  \label{fig:setup}
\end{figure}

\subsection{Overview}

Figure~\ref{fig:setup} illustrates how \name{} measures Web filtering. The
process involves three parties: a Web client that acts as our measurement
vantage point; a measurement target, which is a server that hosts a Web resource
that we suspect is filtered; and an origin Web server, which serves a Web page
to the Web client instructing the client how to collect measurements. In each
page it serves, the origin includes a {\em measurement task}, which is a small
measurement collection program that attempt to access potentially filtered Web
resources (\eg, Web pages, image files) from the target and determine
whether such accesses were successful. The client runs this measurement task
after downloading and rendering the page. The greatest challenge is coping with
browsers' limited APIs for conducting network measurements, particularly when
accessing these resources requires issuing cross-origin requests.

The scope of Web filtering varies in granularity from individual URLs (\eg, a
specific news article or blog post) to entire domains. Detecting Web filtering
is difficult regardless of granularity. On one hand, detection becomes more
difficult with increasing specificity.
%, and any conclusions that we can draw about the technical measurements
%concerning filtering are thus necessarily limited by the scope that we can
%observe.
When specific Web resources are filtered (as opposed to, say, entire domains),
there are fewer ways to detect it. Detecting filtering of entire domains is
relatively straightforward because we have the flexibility to test for such
filtering simply by checking accessibility of a small number of resources hosted
on that domain. In contrast, detecting filtering of a single URL essentially
requires an attempt to access that exact URL. Resource embedding only works with
some types of resources, which further restricts the Web resources we can test
and exacerbates the difficulty of detecting very specific instances of
filtering.

On the other hand, inferring broad filtering is difficult because \name{} can
only observe the accessibility of individual Web resources, and such
observations are binary (\ie, whether or not the resource was reachable). Any
conclusions we draw about the scope of Web filtering must be inferred from
measurements of individual resources. We take a first-order glimpse at such
inferences in Section~\ref{subsec:inference} and present a filtering detection
algorithm in Section~\ref{sec:implementation}.

% Removed this text because it was redundant:
%
%\name{}'s goal is to determine what, where, and when Web resources are filtered;
%we leave determining {\em how} filtering occurs to followup analysis with other
%tools more capable of in-depth analysis from fewer vantage points (\eg,
%OONI~\cite{Filasto:foci2012,www-ooni}). 

%The rest of this section explains how Web clients can use cross-origin
%requests to observe Web filtering of measurement targets.

\subsection{Measurement tasks}\label{sec:measurementTasks}

Measurement tasks are small, self-contained HTML and JavaScript snippets
that attempt to load a Web resource from a measurement
target. \name{}'s measurement tasks must satisfy four requirements:
First, they must be able to successfully load a cross-origin resource in
  locations without Web filtering. Tasks cannot use XMLHttpRequest (\ie, AJAX
  requests), which is the most convenient way to issue cross-origin requests,
  because default Cross-origin Resource Sharing settings prevent such requests
  from loading cross-origin resources from nearly all domains. Instead, we
  induce cross-origin requests by embedding images, style sheets, and scripts
  across domains, which browsers typically allow.

Second, they must provide feedback about whether or not loading a
  cross-origin resource was successful. Several convenient mechanisms
  for loading arbitrary cross-origin requests (\eg, the \texttt{iframe}
  tag) lack a clear way to detect when resources fail to load, and are
  hence unsuitable for measurement tasks.

Third, tasks must not compromise the security of the page running the
  task. Tasks face both client- and server-side security threats. On the
  client side, because \name{} detects Web filtering by {\em embedding}
  content from other origins (rather than simply requesting it, as would
  be possible with an AJAX request), such embedding can pose a threat as
  the browser renders or otherwise evaluates the resource after
  downloading it. In some cases, rendering or evaluating the resources
  is always innocuous (\eg, image files); in other cases (\eg,
  JavaScripts), \name{} must carefully sandbox the embedded
  content to prevent it from affecting other aspects of Web
  browsing. Requesting almost any Web object changes server state, and
  measurement tasks must take these possible side effects into
  account. In some cases, the server simply logs that the request
  happened, but in others, the server might insert rows into a database,
  mutate cookies, change a user's account settings, etc. Although it is
  often impossible to predict such state changes, measurement tasks
  should try to only test URLs without obvious server side-effects.

Finally, measurement tasks must not significantly affect perceived
performance, appearance, or network usage.

Below is an example of a simple measurement task that instructs the Web client
to load an image hosted by a measurement target {\tt censored.com}:

\begin{small}
\begin{verbatim}
<img src="//censored.com/favicon.ico"
     style="display: none"
     onload="submitSuccess()"
     onerror="submitFailure()"/>
\end{verbatim}
\end{small}

\noindent This task meets the four requirements because it (1) requests an image
from a remote measurement target using the {\tt img} tag, which is allowed by
browser security policy; (2) detects whether the browser successfully loaded the
image by listening for the {\tt onload} and {\tt onerror} events; (3) trivially
maintains security by not executing any code from resources served by the
measurement target; and (4) preserves performance and appearance by only loading
a very small icon (typically $16\times16$ pixels) and hiding it using the
\verb+display: none+ style rule. \tr{Section~\ref{sec:system} explains how tasks
submit results using the {\tt submitSuccess} and {\tt submitFailure} functions.}
Appendix~\ref{sec:appendix} presents a longer example of a measurement task.

\subsection{Inferring Web filtering}\label{subsec:inference}

\begin{table*}
\centering
  \begin{tabular}[t]{|l|p{3.2in}|p{2.1in}|}
  \hline
  Mechanism & Summary & Limitations \\ \hline\hline
  Images & Render an image. Browser fires {\tt onload} if successful. & Only small images (\eg, $\leq$ 1~KB). \\ \hline
  Style sheets & Load a style sheet and test its effects. & Only non-empty style sheets. \\ \hline
  %Inline frames (load time) & Record the time taken to render a resource in an iframe. Fast rendering implies page was filtered. & Only detects filtering that fails fast.\newline Only small pages without side effects. \\ \hline
  %Inline frames (caching) & Load a Web page in an iframe, then load an image embedded on that page. Cached images render quickly, implying the page was not filtered. & Only pages with cacheable images.\newline Only small pages (\eg, $\leq$ 100~KB).\newline Only pages without side effects. \\ \hline
  Inline frames & Load a Web page in an iframe, then load an image embedded on that page. Cached images render quickly, implying the page was not filtered. & Only pages with cacheable images.\newline Only small pages (\eg, $\leq$ 100~KB).\newline Only pages without side effects. \\ \hline
  Scripts & Load and evaluate a resource as a script. Chrome fires {\tt onload} iff it fetched the resource with HTTP 200 status. & Only with Chrome.\newline Only with strict MIME type checking. \\ \hline
\end{tabular}
\caption{Measurement tasks use several mechanisms to discover whether Web
resources are filtered. We empirically evaluate parameters for images and inline frames
in Section~\ref{sec:evaluation}.}
\label{tab:tasks}
\end{table*}

A measurement task provides a binary indication of whether a particular resource
failed to load, thus implying filtering of that specific resource. From
collections of these measurements, we can
draw more general conclusions about the scope of filtering, beyond individual
resources (\eg, whether an entire domain is filtered, whether an entire portion
of a Web site is filtered, whether certain keywords are filtered). We must do so
with little additional information about the filtering mechanism. This section
describes how we design sets of measurement tasks to make these inferences.

There are several ways to test accessibility of cross-origin Web resources;
unfortunately, none of them work across all types of filtering, all Web
browsers, and all target sites. Instead, we tailor measurement tasks to each
measurement target and Web client. Detecting Web filtering gets harder as the
scope of filtering becomes more specific, so we start with techniques for
detecting broad-scale filtering and work toward more specific filtering schemes.
Table~\ref{tab:tasks} summarizes the measurement tasks we discuss in this
section. 

\tr{
Section~\ref{sec:evaluation} empirically determines concrete values for
vague claims about requirements for each measurement task (\eg, ``small''
images).
In Section~\ref{sec:implementation}, we show that these measurement tasks can
accurately detect Web filtering, \ie, whether they can load a cross-origin
resource when Web filtering is absent and fail to load the same resource once
filtering is in place. We do so by (1)~emulating several forms of Web filtering
and run controlled experiments, and (2)~verifying that these measurement tasks
detect instances of real Web filtering in several countries.
}

\subsubsection{Filtering of entire domains}

\name{} performs collections of measurement tasks that help infer that a censor
is filtering an entire domain. It is prohibitively expensive to check
accessibility of {\em every} URL hosted on a given domain. Instead, we assume
that if several {\em auxiliary} resources hosted on a domain (\eg, images, style
sheets) are inaccessible, then the entire domain is probably inaccessible.
Our intuition is that rather than filtering many
supporting resources, an adversary will more likely filter the entire domain (or
at least an entire section of a site). Fortunately, detecting filtering of some
auxiliary resources is straightforward because pages often embed them even
across origins. \tr{We discuss two such auxiliary resources.}

\fp{\bf Images.} Web pages commonly embed images, even across origins. Such
embedding is essential for enabling Web services like online advertising and
content distribution networks to serve content across many domains.

\name{} attempts to load and display an image file from a remote origin by
embedding it using the \texttt{<img src=...>} tag. Conveniently, all major
browsers invoke an {\tt onload} event after the browser fetches and renders the
image, and invoke {\tt onerror} if either of those steps fails; the requirement to
successfully render the image means that this mechanism only works for images
files and cannot decide the accessibility of non-image content. Downloading and
rendering an image does not affect user-perceived performance if the image is
small (\eg, an icon), and measurement tasks can easily hide images from view.
This technique only works if the remote origin hosts a small image that
we can
embed.

\fp{\bf Style sheets.} Web pages also commonly load style sheets across origins.
For example, sites often load common style sheets (\eg,
Bootstrap~\cite{www-bootstrap}) from a CDN to boost performance and increase
cache efficiency.

\name{} attempts to load a style sheet using the \texttt{<style src=...>} tag and
detects success by verifying that the browser applied the style
specified by the sheet. For example, if the sheet specifies that the font color
for {\tt <p>} tags is blue, then the task creates a {\tt <p>} tag and checks
whether its color is blue using {\tt getComputedStyle}. To prevent the sheet's
style rules from colliding with those of the parent Web page, we load the sheet
inside an {\tt iframe}. Although some browsers are vulnerable to cross-site
scripting attacks when loading style sheets, these issues have been fixed in all
newer browsers~\cite{huang2010:ccs}. Style sheets are generally small and load
quickly, resulting in negligible performance overhead.

\subsubsection{Filtering of specific Web pages}\label{subsubsec:individual}

Governments sometimes filter one or two Web pages (\eg, blog posts) but leave
the remainder of a domain intact, including resources embedded by the filtered
pages~\cite{nabi2013anatomy}. Detecting this type of filtering is more difficult
because there is less flexibility in the set of resources that \name{} can use
for measurement tasks: it must test accessibility of the Web page in question
and cannot generally determine whether the page is filtered based on the
accessibility of other (possibly related) resources.  Testing filtering of Web
pages, as opposed to individual resources, is significantly more expensive,
complicated, and prone to security vulnerabilities because such testing often
involves fetching not only the page itself, but also fetching all of that page's
referenced objects and rendering everything. This means we must be very careful
in selecting pages to test. Many pages are simply too expensive or open too many
vulnerabilities to test. Section~\ref{sec:system} discusses the infrastructure
and decision process we use to decide whether a Web page is suitable for
testing.

We present two mechanisms for testing Web filtering of Web pages, and the
limitations of each mechanism:

\fp{\bf Inline frames.} A Web page can include any other Web page inside itself
using the {\tt iframe} tag, even across origins. However, browsers place strict
communication barriers between the inline page and the embedding page for
security, and provide no explicit notification about whether an inline frame
loaded successfully.

Instead, the task infers whether the resource loaded successfully by observing
timing. It first attempts to load the page in an iframe; then, after that iframe
finishes load,the task records how long it takes to download and render an image
that was embedded on that page. If rendering this image is fast (\eg, less than
a few milliseconds) we assume that the image was cached from the previous fetch
and therefore the Web page loaded successfully. This approach only
works with pages that embed objects that will be cached by the browser and are
unlikely to have been cached from a prior visit to another Web page; for
example, common images like the Facebook's ``thumbs up'' icon appear on many
pages and may be in the browser cache even if the iframe failed to load. This
approach can be expensive because it requires downloading and rendering entire
Web pages. Additionally, pages can detect when they are rendered in an inline
frame and may block such embedding.

\fp{\bf Scripts.} Web pages often embed scripts across origins, similarly to how
they embed style sheets. For example, many pages embed jQuery and other
JavaScript libraries from a content distribution network or some other
third-party host~\cite{www-jquery}.

The Chrome browser handles script embedding in a way that lets us gauge
accessibility of {\em non-script} resources from a remote origin. Chrome invokes
an {\tt onload} event if it can fetch the resource (\ie, with an {\tt HTTP 200
OK} response), regardless of whether the resource is valid JavaScript. In
particular, Chrome respects the {\tt X-Content-Type-Options: nosniff} header,
which servers use to instruct browsers to prohibit execution of scripts with an
invalid MIME type~\cite{barth2009secure}. Other browsers are not so forgiving, so
we use this task type on Chrome only. This technique is convenient, but it
raises security concerns because other browsers may attempt to execute the
fetched object as JavaScript. Section~\ref{sec:system} describes 
how we make this decision.

\section{\name{} Measurement System}\label{sec:system}

\begin{figure}
  \centering\includegraphics[width=0.9\linewidth]{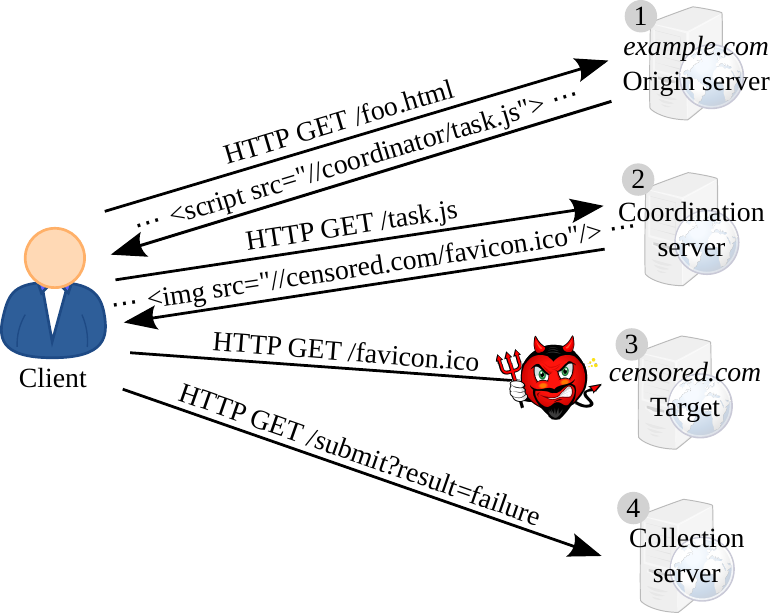}
	\caption{An example of observing Web filtering with \name{}. The
		origin Web page includes \name{}'s measurement script,
		which the coordinator decides should test filtering of
		{\tt censored.com} by attempting to fetch an image. The request
		for this image fails so the client notifies the collection
		server.}
	\label{fig:system}
\end{figure}

This section presents \name{}, a distributed platform for measuring Web
filtering. \name{} selects targets to test for Web filtering
(\S~\ref{sec:sources}), generates measurement tasks to measure those targets
(\S~\ref{sec:generation}), schedules tasks to run on Web clients
(\S~\ref{sec:scheduling}), delivers these tasks to clients for execution
(\S~\ref{sec:distribution}), collects the results of each task
(\S~\ref{sec:collection}), and draws conclusions concerning filtering practices
based on the collective outcomes of these tests using the inference techniques
from Section~\ref{sec:inference}.

Figure~\ref{fig:system} shows an example of how \name{} induces a client to
collect measurements of Web filtering. The client visits a Web site {\tt
http://example.com}, whose webmaster has volunteered to host \name{}. This
origin page references a measurement task hosted on a coordination server; the
client downloads the measurement task, which in turn instructs the client to
attempt to load a resource (\eg, an image) from a measurement target {\tt
censored.com}. This request is filtered, so the client informs a collection
server of this filtering. The remainder of this section explains how the origin
Web server, coordination server, and collection server work together to induce
and collect Web filtering measurements.

\subsection{Sources of measurement targets}\label{sec:sources}

\name{} requires a set of potentially filtered Web sites and resources to test
for Web filtering. This list can contain either specific URLs if
\name{} is testing the reachability of a specific page; or a {\em URL pattern}
denoting sets of URLs (\eg, an entire domain name or URL prefix) to test the
reachability of a domain or a portion of a Web site. A small list of
likely filtered targets
is most useful during initial stages of
deployment when clients of only a few moderately popular Web sites will likely
be contributing measurements. As adoption increases, a broader set of targets
can increase breadth of measurements. We explore how to obtain lists in both
scenarios.

During initial deployment, \name{} relies on third parties to provide lists of
URLs to test for Web filtering. Several organizations maintain such lists. Some
sites rely on per-country experts to curate URLs (\eg, GreatFire for
China~\cite{www-greatfire}, Filbaan for Iran~\cite{www-filbaan}), while others
crowdsource list creation and let anyone contribute reports of Web censorship
(\eg, Herdict~\cite{www-herdict}). Our evaluation in
Section~\ref{sec:evaluation} uses a list of several hundred ``high value'' URLs
curated by Herdict and its partners. Curating accurate and appropriate lists of
potentially censored URLs is beyond the scope of this paper and an active
research area.

If we deploy \name{} to many geographically distributed Web
clients and build a large, accurate Web index, we could instead use \name{}
clients to verify accessibility of the entire Web index, which would avoid the
need for specialized lists of measurement targets by instead testing the entire
Web.
%Section~\ref{sec:evaluation} explores the feasibility and usefulness of
%this deployment mode.
Regardless of whether \name{} curates a small list of
high-value measurement targets or simply extracts URLs from a large Web index,
these URLs and URL patterns serve as input for \name{}'s next stage.

%\sam{This ``blind crawl'' technique only works if we have
%  enough clients in {\em every} country to have a high chance of hitting a
%censored URL. How likely would we be to hit a single censored URL in a small
%country? Are there ways to estimate that? Are there ways to guide/structure our
%Web crawl to make this technique more useful?}

\subsection{Generating measurement tasks}\label{sec:generation}

\begin{figure*}
\centering
\small
\tikzstyle{terminal} = [rectangle, text centered]
\tikzstyle{box} = [draw, rectangle, minimum height=0.4in, text width=0.5in, text centered, line width=1pt]
\tikzstyle{edge} = [text width=0.4in, text centered]
\tikzstyle{line} = [draw, -latex, line width=1pt]
\begin{tikzpicture}[node distance=1.4in, auto]
  \node [terminal, text width=1in] (init) {Measurement target list (\S\ref{sec:sources})};
  \node [box, right of=init] (expand) {Pattern Expander};
  \node [box, right of=expand] (fetch) {Target Fetcher};
  \node [box, right of=fetch] (generate) {Task Generator};
  \node [terminal, text width=1in, right of=generate] (dist) {Task scheduling\\ (\S\ref{sec:scheduling})};

  \path [line] (init) -- node [edge] {Patterns} (expand);
  \path [line] (expand) -- node [edge] {URLs} (fetch);
  \path [line] (fetch) -- node [edge] {HARs} (generate);
  \path [line] (generate) -- node [edge] {Tasks} (dist);
\end{tikzpicture}
\caption{\name{} transforms a list of URL patterns to a set of
measurement tasks in three steps. A URL pattern denotes a set of URL (\eg, all
URLs on a domain). A HAR is an HTTP Archive~\cite{www-har-spec}.}
\label{fig:generation}
\end{figure*}
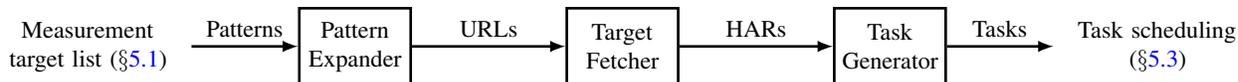

Measurement task generation is a three-step procedure that transforms URL
patterns from the list of measurement targets into a set of measurement tasks
that can determine whether the resources denoted by those URL patterns are
filtered for a client. This procedure happens prior to interaction with clients
(\eg, once per day). Figure~\ref{fig:generation} summarizes the process. First,
the {\em Pattern Expander} transforms each URL pattern into a set of URLs by
searching for URLs on the Web that match the pattern. Second, the {\em Target
Fetcher} collects detailed information about each URL by loading and rendering
it in a real Web browser and recording its behavior in an HTTP Archive (HAR)
file~\cite{www-har-spec}. Finally, the {\em Task Generator} examines each HAR
file to determine which of \name{}'s measurement task types, if any, can measure
each resource and generates measurement tasks for that subset of resources.

The Pattern Expander searches for URLs that match each URL pattern. This step
identifies a set of URLs that can all indicate reachability of a single
resource; for example, all URLs with the prefix {\tt http://foo.com/} are
candidates for detecting filtering of the {\tt foo.com} domain. Some patterns
are trivial (\ie, they match a single URL) and require no work. The rest
require discovering URLs that match the pattern. We currently expand URL
patterns to a sample of up to 50 URLs by scraping site-specific results (\ie,
using the {\em site:} search operator) from a popular search engine. In the
future, \name{} could use its own Web crawler to explore each pattern.

After expanding URL patterns into a larger set of URLs, the Target Fetcher
renders each URL in a Web browser and records a HAR file, which documents the
set of resources that a browser downloads while rendering a URL, timing
information for each operation, and the HTTP headers of each request and
response, among other metadata. We use the PhantomJS~\cite{www-phantomjs}
headless browser hosted on servers at Georgia Tech. To the best of our
knowledge, Georgia Tech does not filter Web requests, especially to the set of
URLs we consider in this paper.

Finally, the Task Generator analyzes each HAR file to determine which subset of
resources is suitable for measuring using one of the types of measurement tasks
from Table~\ref{tab:tasks}. It examines timing and network usage of each
resource to decide whether a resource is small enough to load from an origin
server without significantly affecting user experience, then inspects content
type and caching headers to determine whether a resource matches one of the
measurement tasks. The Task Generator is particularly conservative when
considering inline frames because loading full Web pages can severely impact
performance and user experience (\eg, by playing music or videos). Our prototype
implementation excludes pages that load flash applets, videos, or any other
large objects totaling more than 100~KB, and requires manual verification of
pages before deployment; a future implementation could apply stricter controls.
Refer back to Section~\ref{sec:inference} for more information on the
requirements for each type of measurement task. Section~\ref{sec:study} further
explores network overhead of measurement tasks.

\subsection{Scheduling measurement tasks}\label{sec:scheduling}

After generating measurement tasks, the coordination server must decide which
task to schedule on each client. Task scheduling serves two purposes. First, it
enables clients to run measurements that meet their restrictions. For example,
we should only schedule the script task type from Table~\ref{tab:tasks} on
clients running Chrome. In other cases, we may wish to schedule additional tasks
on clients that remain idle on an origin Web page for a long time. Second,
intelligent task scheduling enables \name{} to move beyond analyzing individual
measurements and draw conclusions by comparing measurements between clients,
countries, and ISPs. For example, a single client in Pakistan could report
failure to access a URL for a variety of reasons other than Web filtering (\eg,
high client system load, transient DNS failure, WiFi unreliability).
However, if 100 clients measure the same URL within 60 seconds of each other and
the only clients that report failure are 10 clients in Pakistan, then we can
draw much stronger conclusions about the presence of Web filtering.

%Our prototype implementation enforces client restrictions but otherwise
%schedules tasks completely randomly. Future versions of \name{} will schedule
%tasks in a rounds 

\subsection{Delivering measurement tasks}\label{sec:distribution}

After scheduling measurement tasks for execution, \name{} must deliver tasks to
these clients, who subsequently run them and issue cross-origin requests for
potentially filtered Web resources. To collect a significant number of useful
Web filtering measurements, \name{} requires a large client population that is
likely to experience a diversity of Web filtering. Previous censorship
measurement efforts require researchers to recruit vantage points individually
and instruct them to install custom software, which presents a significant
deployment barrier~\cite{Filasto:foci2012,www-oni}. In contrast, \name{}
recruits a relatively small number of webmasters and piggybacks on their
sites' existing Web traffic, instantly enlisting nearly {\em all} of these
sites' visitors as measurement collection agents.
%The main deployment challenge for \name{} is instead recruiting webmasters to
%install \name{}'s measurement scripts, ensuring these scripts do not degrade
%user experience, and preventing the scripts from being blocked.

A webmaster can enable \name{} in several ways. The simplest method is to add a
single {\tt <iframe>} tag that directs clients to load an external JavaScript
directly from the coordination server. The coordination server generates a
measurement task specific to the client on-the-fly. This method is attractive
because it requires no server-side modifications, aside from a single tag;
incurs little server overhead (\ie, only the extra time and space required to
transmit that single line); and allows the coordination server to tailor
measurement tasks to individual clients in real time. Unfortunately, this method
is also easiest for censors to fingerprint and disrupt: a censor can simply
block access to the coordination server, which inflicts no collateral damage.
Section~\ref{sec:security} discusses ways to make task delivery more robust to
blocking, while Section~\ref{sec:incentives} discusses incentives for webmasters
to include \name{} on their sites in the first place.

Rather than recruit webmasters ourselves, we have explored the possibility of
purchasing online advertisements and delivering \name{} measurement tasks inside
them. This idea is attractive because online advertising networks already have
established agreements with webmasters to display content (\ie, by paying
webmasters to display ads.) Ad networks even allow advertisers to target ads to
specific users, which \name{} could leverage to measure censorship in specific
countries. Unfortunately for us, this idea works poorly in practice because most
ad networks prevent advertisements from running custom JavaScript and loading
resources from remote origins, with good reason; only a few niche ad networks
are capable of hosting \name{}. Even if more networks could serve \name{}
measurement tasks, they may not take kindly to perceived misuse of their
service, especially if it leads to network filtering and subsequent loss of
revenue in countries wishing to suppress \name{}'s measurements.

%Section~\ref{sec:discussion} discusses the ethics of deploying \name{}'s
%measurement scripts.

%   - Web owner "donates" a small space on his page to help measure censorship.
%     - Either directly (e.g., load a script) or indirectly (e.g., ad space)
%     - Incentives:
%       - Goodwill, transparency.
%       - Reciprocity. We could offer to add the owner's site to the set of
%         targets in exchange for hosting our measurements on his site. This could
%         give feedback on whether the site is available in other countries; this could be
%         especially useful to larger sites, as a crowdsourced monitoring feature.
%     - Cost to owner is negligible, as discussed later.

%   - Unwitting Web user loads the Web page.
%     - Page either contains a direct link to content to test, or a link to
%       script which will in turn link to the content to test.
%     - Javascript obfuscation could help disguise the script en route to user. e.g., FTE.
%     - Various factors can cause this to fail; discuss elsewhere.
%     - There might be ethical concerns here; see discussion section.

\subsection{Collecting measurement results}\label{sec:collection}

After clients run a measurement task, they submit the result of the task for
analysis. Clients submit the result of task (\ie, whether the client could
successfully load the cross-origin resource), related timing information (\ie,
how long it took to load the resource), and the task's measurement ID. The
process of submitting results is similar to the process that clients use to
obtain measurement tasks. In the absence of interference from the adversary,
clients submit results by issuing an AJAX request containing the results
directly to our collection server. Section~\ref{sec:security} discusses
other ways to submit results if the adversary filters access to the
collection server.

\section{Feasibility of \name{} Deployment}\label{sec:evaluation}

We evaluate the feasibility of deploying \name{} based on early experience with
a prototype implementation and analysis of potential measurement
targets. 
This
section addresses three questions about \name{}'s deployment: 
(1)~whether real Web sites are amenable to filtering detection using \name{}'s
measurement tasks, which we explore with offline analysis of
potentially-filtered Web sites;
(2)~whether users visit origin sites, run measurement tasks, and collect
measurements, which we estimate using analytics data collected from a likely
site of \name{} deployment;
(3)~if webmasters will install \name{}, which we study in terms of webmaster
incentives and estimated deployment costs.

\subsection{Are sites amenable to \name{}'s tasks?}\label{sec:study}

This section investigates whether real Web sites host resources that \name{}'s
measurement tasks can use to measure filtering.  We evaluate the feasibility of
using \name{} to measure filtering of both entire domain names and individual
URLs.  To measure filtering practices, we use a list of domains and URLs that
are ``high value'' for censorship measurement according to Herdict and its
partners~\cite{www-herdict-lists}; most sites are either perceived as likely
filtering targets in many countries (\eg, because they are affiliated with human
rights and press freedom organizations) or would cause substantial disruption if
filtered (\eg, social media like Twitter and YouTube). This list contains over
200 URL patterns, of which only 178 were online when we performed our feasibility analysis
in February 2014.

We collect data for this set of experiments by running the first two stages of
the pipeline in Figure~\ref{fig:generation}, which uses the Pattern Expander to
generate a list of 6,548 URLs from the 178 URL patterns in our list, then
collect HAR files for each URL using the Target Fetcher. We then send these HAR
files to a modified version of the Task Generator that emits statistics about
sizes of accepted resources and pages.

\paragraph{Filtering of entire domains.} We explore whether \name{} can measure
filtering of each of the 178 domains on the list we generated as
described above. Recall from
Section~\ref{subsec:inference} that we can use either images or style sheets to
observe Web filtering of an entire domain; for simplicity, this analysis only
considers images, although style sheets work similarly. We can measure a domain
using this technique if (1)~it contains images that can be embedded by an origin
site and (2)~those images are small enough not to significantly affect user
experience. We explore both of these requirements for the 178 domains in our
list. Because our implementation expands URL patterns using the top 50
search results for that pattern, we will be analyzing a sample of at most 50
URLs per domain. Most of these domains have more than 50 pages, so our results
are a lower bound of the amenability of \name{} to collect censorship
measurements from each domain.

\begin{figure}[t]
  \includegraphics{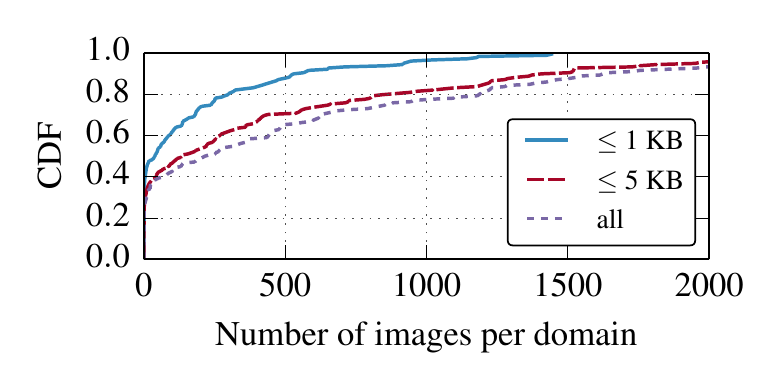}
  \caption{Distribution of the number of images hosted by each of the 178
    domains tested, for images that are at most 1~KB, at most 5~KB, and any
    size. Over 60\% of domains host images that could be delivered to clients
    inside a single packet, and a third of domains have hundreds of such images
    to choose from.}
  \label{fig:imagesPerDomain}
\end{figure}

Figure~\ref{fig:imagesPerDomain} plots the distribution of the number of images
that each domain hosts. 70\% of domains embed at least one image, and almost all
such images are less than 5~KB. Nearly as many domains embed images that fit
within a single packet, and a third of domains have hundreds of such images. Even
if we conservatively restrict measurement tasks to load images less than 1~KB,
\name{} can measure Web filtering of over half of the domains.

\paragraph{Filtering of specific Web pages.} We explore how often \name{} can
measure filtering of individual URLs \tr{hosted on the domains from the list we
constructed. Section~\ref{subsubsec:individual} explained how to do this using
inline frames or browser-specific script embedding. We explore the feasibility
of the first, which attempts to} by loading a Web page in an iframe and verifying that
the browser cached embedded resources from that page. We can use this mechanism
to measure filtering of pages that (1) do not incur too much network overhead
when loading in a hidden iframe and (2) embed cacheable images. \tr{Analysis of the
``inline frames (load time)'' mechanism requires complex parameter tuning that
requires a lot of timing from a diverse client population. Although the ``script
embedding'' mechanism works with nearly all URLs, the technique only works for
users who are browsing using Chrome.}

\begin{figure}[t]
  \includegraphics{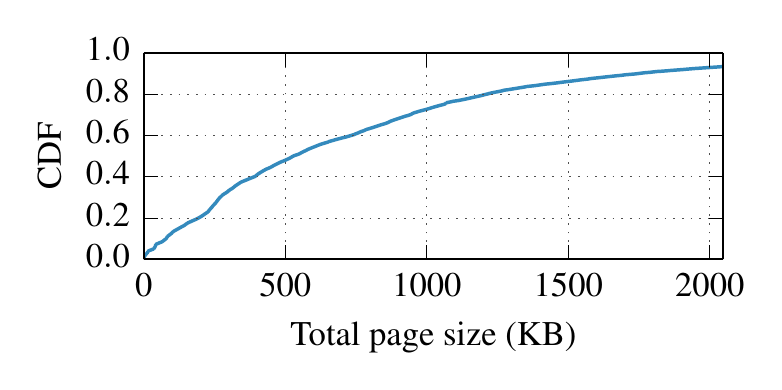}
  \caption{Distribution of page sizes, computed as the sum of sizes of all
  objects loaded by a page. This indicates the network overhead each page would
  incur if a measurement task loaded it in a hidden iframe. Over half of pages
  load at least half a megabyte of objects.}
  \label{fig:pageSizes}
\end{figure}

We first study the expected network overhead from loading sites in an iframe.
Figure~\ref{fig:pageSizes} plots the distribution of page sizes for each URL,
where the page size is the sum of sizes of all resources a page loads and is a
rough lower bound on the network overhead that would be incurred by loading each
page in a hidden iframe (protocol negotiation and inefficiencies add further
overhead). Page sizes are distributed relatively evenly between 0--2~MB with a
very long tail. Our prototype only permits measurement tasks to load pages
smaller than 100~KB, although future implementations might tune this bound to a
client's performance and preferences.

We then evaluate whether these sites embed content that can be retrieved with
cross-origin requests. Figure~\ref{fig:cacheable} shows the distribution of the
number of cacheable images per URL for pages that are at most 100~KB, at most
500~KB, and any size. Nearly 70\% of pages embed at least one cacheable image
and half of pages cache five or more images, but these numbers drop
significantly when restricting page sizes. Only 30\% of pages that are at most
100~KB embed at least one cacheable image.

\name{} can measure filtering of upwards of 50\% of domains depending on the
sizes of images, but fewer than 10\% of URLs when we limit pages to 100~KB. This
finding supports our earlier observation in Section~\ref{subsec:inference} that
detecting the filtering of individual Web resources may be significantly more
difficult than detecting the filtering of entire domains.

\begin{figure}
  \includegraphics{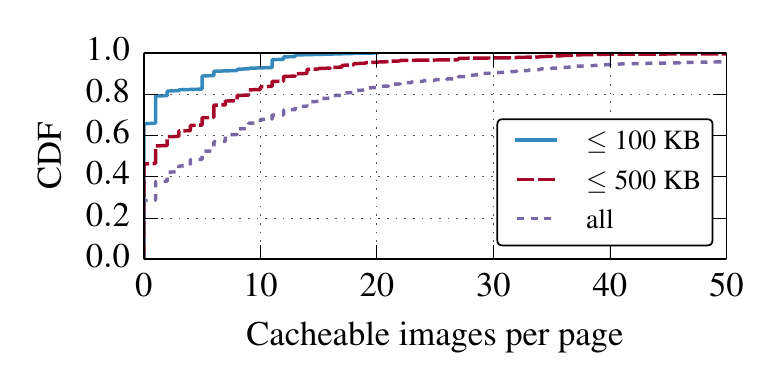}
  \caption{Distribution of the number of cacheable images loaded by pages that
    require at most 100~KB of traffic to load, pages that incur at most 500~KB of traffic,
    and all pages. Perhaps unsurprisingly, smaller pages contain fewer
    (cacheable) images. Over 70\% of all pages cache at least one image and half
  of all pages cache five or more images; these numbers drop considerably when
excluding pages greater than 100~KB.}
  \label{fig:cacheable}
\end{figure}

\subsection{Who performs \name{} measurements? }\label{sec:pilot}

\name{} requires clients to visit the Web sites that are hosting \name{}
scripts.  The demographics of clients who perform \name{} measurements is
closely related those who visit a participating Web
site. To evaluate whether a typical Web site will receive measurements from
enough locations, we examined demographic data collected by Google Analytics for
the home page of a professor in February 2014~\cite{www-google-analytics}.

The site saw $1,171$ visits during course of the month. Most visitors were from
the United States, but we saw more than 10 users from 10 other countries, and
16\% of visitors reside in countries with well-known Web filtering policies
(India, China, Pakistan, the UK, and South Korea), indicating that dispatching
measurement tasks to sites such as academic Web pages may yield measurements
from a variety of representative locations. Of these visitors, $999$ attempted
to run a measurement task; we confirmed nearly all of the rest to be automated
traffic from our campus' security scanner. We also found that 45\% of visitors
remained on the page for longer than 10 seconds, which is more than sufficient
time to execute at least one measurement task and report its results. The 35\%
of visitors who remained for longer than a minute could easily run multiple
measurement tasks.

Our small pilot deployment of \name{} is representative of the sites
where we can expect \name{} to be deployed in the short term. Although
adoption of \name{} by even a single high-traffic Web site would
entirely eclipse measurements collected by these small university
deployments, grassroots recruitment remains necessary: \name{} relies on
a variety of origin sites to deter an adversary from simply blocking
access to all origins to suppress our measurement collection.
Section~\ref{sec:security} discusses further mechanisms for deterring
filtering of \name{}'s origin sites and backend infrastructure.

\subsection{Will webmasters install \name{}?}\label{sec:incentives}

\name{} cannot directly target specific demographics for measurement
collection---the measurements that we collect arise from the set of users who
happen to visit a Web site that has installed an \name{} script. If the sites
that host \name{} are globally popular (\eg, Google), then \name{} can achieve
an extremely widespread sampling of users;  on the other hand, if the sites are
only popular in particular regions, the resulting measurements will be limited
to those regions.
%The best solution to this problem is to achieve deployment across sites that
%are likely to be popular among client populations where we would like
%measurements or to deploy on globally popular sites.

Recruiting webmasters to include \name{}'s measurement scripts should be
feasible. First, installing \name{} on a Web site incurs little cost. Serving
these scripts to clients adds minimal network overhead; our prototype adds only
100 bytes to each origin page and requires no additional requests or connections
between the client and the origin server. Measurements themselves have little
effect on the Web page's perceived performance because they run asynchronously
after the page has loaded and rendered. However, they do incur some network
overhead to clients when loading cross-origin resources, which may be
undesirable to users with bandwidth caps or slow, shared network connections. As
Section~\ref{sec:study} explained, measurement tasks that detect filtering of a
domain (\ie, by loading small images) incur overheads that are usually an
insignificant fraction of a page's network usage.

Second, we see two strong incentives for webmasters to participate in \name{}.
Many webmasters may support \name{} simply out of greater interest in measuring
Web filtering and encouraging transparency of government censorship. The
grassroots success of similar online freedom projects (\eg,
Tor~\cite{Dingledine2004}) in recruiting 
volunteers to host relays and bridges suggests that such a population does exist.
For further incentive, we could institute a reciprocity agreement for
webmasters: in exchange for installing our measurement scripts, webmasters could
add their own site to \name{}'s list of targets and receive notification about
their site's availability from \name{}'s client population.

\section{Measurements}\label{sec:implementation}

We confirm the soundness of \name{}'s measurement tasks with
both controlled experiments and by comparing our ability to confirm cases of Web
filtering with independent reports of filtering from other research studies.  We
have implemented and released every component of \name{} described in
Section~\ref{sec:system} and have collected seven months of measurements from
May~2014 through January~2015. %~\cite{www-encore-github}.  Removed for blind review.
%Early results from this deployment have confirmed several known cases of Web
%filtering.

To date, at least 17 volunteers have deployed \name{} on their sites, although
the true number is probably much higher; $\nicefrac{3}{4}$ of measurements come from
sites that elect to strip the {\tt Referer:} header when sending results. We
recorded 141,626 measurements from 88,260 distinct IPs in 170 countries, with
China, India, the United Kingdom, and Brazil reporting at least 1,000
measurements, and more than 100 measurements from Egypt, South Korea, Iran,
Pakistan, Turkey, and Saudi Arabia. These countries practice some form of Web
filtering. We use a standard IP geolocation database to determine client
locations~\cite{www-geoip}. Clients ran a variety of Web browsers and operating systems.

% - Basic statistics:
%   - How many sites
%   - Number of visitors
%   - Visitors per county

\subsection{Are measurement tasks sound?}

To confirm the soundness of \name{}'s measurements, we built a Web censorship
testbed, which has DNS, firewall, and Web server configurations that emulate
seven varieties of DNS, IP, and HTTP filtering.
%~\cite{www-encore-testbed}.  Removed for blind review.
For
three months, we instructed approximately 30\% of clients to measure resources
hosted by the testbed (or unfiltered control resources) using the four task
types from Table~\ref{tab:tasks}.
%To minimize the possibility of drawing false conclusions in the unlikely event
%that the testbed were blocked in some regions, we ran the testbed on hosts in
%the same domain as the \name{} coordination and collection server. We used the
%testbed to verify that can report filtering for seven varieties of DNS, IP, and
%HTTP filtering and report no filtering for unfiltered resources.
For example, we verified that the {\em images} task type detects DNS blocking by
attempting to load an image from an invalid domain and observing that the task
reports filtering; we verified that the same task successfully loads an
unfiltered image.
%\footnote{From {\tt //invalid.noise.gatech.edu/pixel.png} and {\tt
%//encore.noise.gatech.edu:8892/pixel.png}, respectively.}

Verification is straightforward for the image, style sheet, and script task
types because they give explicit binary feedback about whether a resource
successfully loaded. \name{} collected 8,573 measurements for these task
types; after excluding erroneously contributed measurements (\eg, from Web
crawlers), there were no true positives and few false positives. For
example, clients in India, a country with notoriously unreliable network
connectivity, contributed to a 5\% false positive rate for images.

Verifying soundness of the inline frame task type requires more care because it
infers existence of filtering from the time taken to load resources.
Figure~\ref{fig:iframeCache} compares the time taken to load an uncached versus
cached single pixel image from 1,099 globally distributed \name{} clients.
Cached images normally load within a few tens of milliseconds, whereas most
clients take at least 50~ms longer to load the same image uncached. The few
clients with little difference between cached and uncached load time were
located on the same local network as the server. Difference in load time
will be more pronounced for larger images and with greater latency between
clients and content.

In both cases, false positives highlight (1) that distinguishing Web filtering
from other kinds of network problems is difficult and (2) the importance of
collecting many measurements before drawing strong conclusions about Web
filtering. We now develop a filtering detection algorithm that addresses both
concerns.

\begin{figure}
\includegraphics{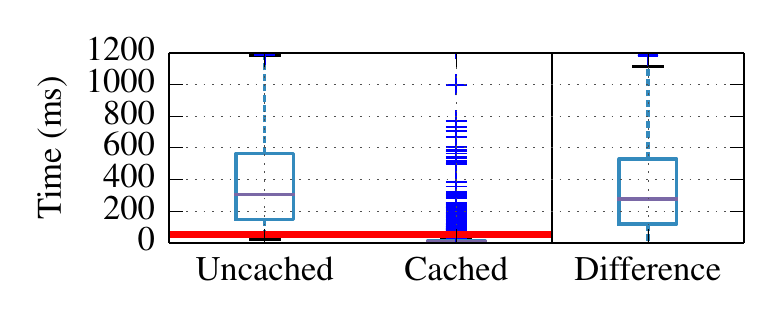}
\caption{Comparison between load times for cached and uncached images from 1,099
  \name{} clients. Cached images typically load within tens of milliseconds,
  whereas uncached usually take at least 50~ms longer to load, indicated by the
  bold red line. We use this difference to infer filtering.}\label{fig:iframeCache}
\end{figure}

%   - Perform control measurements
%     - Why? Confirm soundness of our technique.
%       - Measuring against the same server minimizes risk of false conclusions.
%     - Which ones?
%       - NxM set of experiments
%       - Control servers
%       - Measurement types
%     - Do they work?

\subsection{Does \name{} detect Web filtering?}\label{sec:implementation:detection}

% - Measurements can fail for many reasons
% - What do we want to detect?
% - Our current detection criteria
% - Results
% - Areas for improvement

We instructed the remaining 70\% of clients to measure resources suspected of
filtering, with the goal of independently verifying Web filtering reported in
prior work. Because measuring Web filtering may place some users at risk, we
only measured Facebook, YouTube, and Twitter. These sites pose little additional
risk to users because browsers already routinely contact them via cross-origin
requests without user consent (\eg, the Facebook ``thumbs up'' button, embedded
YouTube videos and Twitter feeds). Expanding our measurements to less popular
sites would require extra care, as we discuss in the next section.

We aimed to detect resources that are consistently inaccessible from one region,
yet still accessible from others. For this purpose, we measure filtering of
entire domains, using the image task type. This is challenging because
measurement tasks may fail for reasons other than filtering: clients may
experience intermittent network connectivity problems, browsers may incorrectly
execute measurement tasks, sites may themselves go offline, and so on. We use a
statistical hypothesis test to distinguish such sporadic or localized
measurement failures from more consistent failures that might indicate Web
filtering. We model each measurement success as a Bernoulli random variable with
parameter $p = 0.7$; we assume that, in the absence of filtering, clients should
successfully load resources at least 70\% of the time. Although this assumption
is conservative, it captures our desire to eliminate false positives, which can
easily drown out true positives when detecting rare events like Web filtering.
For each resource and region, we count both the total number of measurements
$n_r$ and the number of successful measurements $x_r$ and run a one-sided
hypothesis test for a binomial distribution; we consider a resource as filtered
in region $r$ if $x_r$ fails this test at 0.05 significance (\ie,
$\mathrm{Pr}[\mathsf{Binomial}(n_r, p) \leq x_r] \leq 0.05$) yet does not fail
the same test in other regions.

Applying this technique on preliminary measurements confirms well-known
censorship of {\tt youtube.com} in Pakistan, Iran, and
China~\cite{www-transparencyreport}, and of {\tt twitter.com} and {\tt
facebook.com} in China and Iran.
%Measurements also appear to confirm Turkey's widely-publicized filtering of
%YouTube at the end of March~\cite{www.youtube-turkey-2014}, but we didn't
%collect enough measurements for statistical significance.
%We are actively recruiting more sites and collecting more measurements, and
%continue to expand these results.
Although our detection algorithm works well on
preliminary data, possible enhancements include dynamically tuning model
parameters to account for differing false positive rates in each country and
accounting for potential confounding factors like user behavior differences
between browsers and ISPs~\cite{tariq:nano:conext2009}.

% - Purpose of deployment
%   - Measure real sites
%     - Which ones?
%     - Which types of measurements?
%     - Any interesting results?

\section{Ethics and Security}\label{sec:security}

This section discusses barriers to \name{}'s widespread deployment, from the
ethics of collecting measurements from unsuspecting Web users to the potential
for attackers to block, disrupt, or tamper with client measurements or
collection infrastructure.

%We discuss several tactics that can mitigate attacks, but
%\name{} ultimately faces fundamental security limitations because it operates
%within Web browsers, which have limited security APIs and may themselves have
%security vulnerabilities. In particular, because \name{} cannot establish trust
%of clients performing measurements, it can do little to prevent an adversary
%from contributing false measurements; any system accepting measurements from
%anonymous or pseudonymous clients faces this risk.

\begin{table*}
\footnotesize
\centering
  \begin{tabular}{|l|p{4.8in}|}\hline
    Date & Event \\ \hline\hline
    February 2014 and prior & Informal discussions with Georgia Tech IRB conclude that \name{} (and similar work) is not human subjects research and does not merit formal IRB review. \\
    % February 28, 2014 & First paper submission, without measurement results. \\
    March 13, 2014 -- March 24, 2014 & \name{} begins collecting measurements from real users using a list of over 300 URLs. We're unsure of the exact date when collection began because of data loss. \\
    March 18, 2014 & We begin discussing \name{}'s ethics with a researcher at the Oxford Internet Institute. \\
    April 2, 2014 & To combat data sparsity, we configure \name{} to only measure favicons~\cite{www-w3-favicons}. The URLs we removed were a subset of those we crawled from \S\ref{sec:generation}. \\
    May 5, 2014 & Out of ethical concern, we restrict \name{} to measure favicons on only a few sites. \\
    May 7, 2014 & Submission to IMC 2014, which includes results derived from our March 13 URL list. \\
    September 17, 2014 & Georgia Tech IRB officially declines to review \name{}. We requested this review in response to skeptical feedback from IMC. \\
    September 25, 2014 & Submission to NSDI 2015, using our URL list on April 2. \\
    January 30, 2015 & Submission to SIGCOMM 2015, using our URL list on May 5. \\
    February 6, 2015 & Princeton IRB reaffirms that \name{} is not human subjects research. We sought this review at the request of the SIGCOMM PC chairs after Nick Feamster moved to Princeton. \\ \hline
  \end{tabular}
  \caption{Timeline of \name{} measurement collection, ethics discussions, and
    paper submissions. As our understanding of \name{}'s ethical implications
    evolved, we increasingly restricted the set of measurements we collect and
    report. See \url{http://encore.noise.gatech.edu/urls.html} for information on how
    the set of URLs that \name{} measures has evolved over time.}
  \label{tab:milestones}
\end{table*}

\paragraph{Which resources are safe to measure?} \name{} induces clients to
request URLs that might be incriminating in some countries and circumstances.
In particular, the most interesting URLs to measure may be those most likely to
get users into trouble for measuring them. Curating a list of target URLs
requires striking a balance between ubiquitous yet uninteresting URLs (\eg,
online advertisers, Google Analytics, Facebook) and obscure URLs that
governments are likely to censor (\eg, human rights groups). Although our
work does not prescribe a specific use case, we recognize that deploying a tool
like \name{} engenders risks that we need to better understand.

Balancing the benefit and risk of measuring filtering with \name{} is difficult.
This paper has made the benefit clear: \name{} enables researchers to collect
new data about filtering from a diversity of vantage points that was previously
prohibitively expensive to obtain and coordinate. Ongoing efforts to measure Web
filtering would benefit from \name{}'s diversity and systematic
rigor~\cite{www-oni,www-centinel,Filasto:foci2012}. The risk that \name{} poses
are far more nebulous: laws against accessing filtered content vary from country
to country, and may be effectively unenforceable given the ease with which sites
(like \name{}) can request cross-origin resources without consent; there is no
ground truth about the legal and safety risks posed by collecting network
measurements.

Striking this balance between benefit and risk raises ethical questions that
researchers in computer science rarely face and that conventional ethical
standards do not address. As such, our understanding of the ethical implications
of collecting measurements using \name{} has evolved, and the set of
measurements we collect and report on has likewise changed to reflect our
understanding. Table~\ref{tab:milestones} highlights a few key milestones in
\name{}'s deployment, which has culminated in the set of measurements we report
on in this paper. The Institutional Review Boards (IRBs) at both Georgia Tech
and Princeton declined to formally review \name{} because it does not collect or
analyze Personally Identifiable Information (PII) and is not human subjects
research~\cite{irb-email}. Yet, \name{} is clearly capable of exposing its users
to some level of risk. Because we do not understand the risks that a tool like
\name{} presents, we have focused most of our research efforts on developing the
measurement technology, not on reporting results from the measurements that we
gather.  Other censorship measurement tools have and will continue to face
similar ethical questions, and we believe that our role as researchers is to
lead a responsible dialogue in the context of these emerging tools.

As part of this ongoing dialogue, we hope that the community will develop ethical
norms that are grounded in theory, applicable in practice, and informed by
experts. To this end, we have discussed \name{} with ethics experts at the
Oxford Internet Institute, the Berkman Center, and Citizen Lab, and our follow
on work examines broader ethical concerns of censorship
measurement~\cite{Jones2015Ethics}. We have also been working with the
organizers of the SIGCOMM NS Ethics workshop~\cite{www-workshop-ns-ethics15},
which we helped solicit, to ensure that its attendees will gain experience
applying principled ethical frameworks to networking and systems research, a
process we hope will result in more informed and grounded discussions of ethics
in our community.

Schechter~\cite{schechter14} surveyed people about the ethics of various
research studies, including \name{}, and found that most people felt that
unconstrained use of \name{} would be highly unethical. However, as the report
acknowledges, the survey didn't elaborate on the inherent risks of browsing {\em
any} Web page, the potential benefits of research like \name{}, the risks of
alternative means of measuring censorship, or low-risk deployment modes.

\name{} underscores the need for stricter cross-origin security
policy~\cite{www-content-security-policy}. Our work exploits existing
weaknesses, and if these policies could endanger users then strengthening those
policies is clearly a problem worthy of further research.

\fp{\bf Why not informed consent?} The question of whether to obtain informed
consent is more complicated than it might first appear. Informed consent is not
always appropriate, given that in disciplines where experimental protocols for
human subjects research are well-established, there are classes of experiments
that can still be conducted ethically without it, such as when obtaining consent
is either prohibitive or impractical and there is little appreciable risk of
harm to the subject.

Researchers and engineers who have performed large-scale network measurements
can appreciate that obtaining consent of any kind is typically impractical. For
\name{}, it would require apprising a user about nuanced technical concepts,
such as cross-origin requests and how Web trackers work---and doing so across
language barriers, no less. Such burdens would dramatically reduce the scale and
scope of measurements, relegating us to the already extremely dangerous status
quo of activists and researchers who put themselves into harm's way to study
censorship. Even if we could somehow obtain consent at scale, informed consent
does not {\em ever} decrease risk to users; it only alleviates researchers from
some responsibility for that risk, and may even increase risk to users by
removing any traces of plausible deniability.

We believe researchers should instead focus on reducing risk to uninformed
users, as we have done with repeated iteration after consultation with ethics
experts. It is generally accepted that users already have little control over or
knowledge of much of the traffic that their Web browsers and devices generate (a
point raised by Princeton's office of research integrity and assurance), which
already gives users reasonable cover. By analogy, the prevalence of malware and
third-party trackers itself lends credibility to the argument that a user cannot
reasonably control the traffic that their devices send. The more widespread
measurements like \name{} become, the less risky they are for users.

%consent may even result in unintended outcomes that
%actually increase risk to users. 

%It is also worth noting that informed consent does not {\em ever}
%decrease risk to users and may sometimes increase that risk.
%Asking users to
%explicitly acknowledge that they are performing such measurements---particularly
%when they are as innocuous as retrieving favicons and content that browsers may
%be retrieving anyway without the user's knowledge---may in fact expose them to
%additional risk.

%Although our
%conversations with citizens of various countries suggest that attempting to
%access censored content is not illegal {\em per se}, simply because an activity
%is legal does not make it harmless, so any deployment of \name{} must be
%sensitive to inducing users to retrieve content that might land them in trouble.
%Because \name{}'s measurements are effectively a side effect of a user visiting
%an unrelated host site, every client has plausible deniability; most users are
%unaware of \name{}, which runs in the background and has no visible effects.
%\name{} could also shield clients from more incriminating content---for example,
%loading a small icon from a banned site may be less incriminating than loading
%actual content.
%Another concern is that fetching cross-origin requests requires the client to
%retrieve more content than it otherwise would, which is a significant concern
%in countries where data is expensive.

\paragraph{Filtering access to \name{} infrastructure.} Clients can only use
\name{} if they can fetch a measurement task. If the domain (or URL) that hosts
measurement tasks is itself blocked, clients will not be able to execute
measurements. Once a client retrieves a measurement task, subsequent requests
appear as ordinary cross-origin requests; as a result, the main concern is ensuring
that clients can retrieve measurement tasks in the first place.

The server that dispatches tasks could be replicated across many domains
to make it more difficult for a censor to block \name{} by censoring a
single domain.  Clients could contact the coordination server indirectly
via an intermediary or create mirrors of the coordination server
in shared hosting environments (\eg, Amazon AWS),
thereby increasing the collateral damage of blocking a mirror. Going
further, webmasters could contact the coordination server on behalf of
clients (\eg, with a WordPress plugin or Django package) by querying the
coordination server and including the returned measurement task directly
in the page it serves to the client; to increase scalability and
decrease latency, servers could cache several tasks in advance.
Similarly, collection of the results could be distributed across servers
hosted in different domains, to ensure that collection is not blocked.

There are limits to \name{}'s ability to withstand such attacks. Because it runs
entirely within a Web browser, \name{} cannot leverage stronger security tools
like Tor to anonymously report measurements~\cite{Winter2013towards, Dingledine2004}.

\paragraph{Detecting and interfering with \name{} measurements.} Blocking
\name{} based on the contents of measurement tasks (\eg, via deep packet
inspection) should be difficult, because we can easily disguise tasks'
code using JavaScript obfuscation or detection evasion
techniques~\cite{howard2010malware, dyer2013}. Identifying task behavior is
equally difficult because it appears merely as requests to load a cross-origin
object---something many Web sites do under normal operation.  If a single client
performs a {\em sequence} of cross-origin requests that appear unrelated to the
content of the host site, a censor may recognize the sequence as unusual and
either block the subsequent reports or otherwise attempt to distort the results.
We expect such interference to be relatively difficult, however, since a censor
would first have to identify a sequence of requests as a measurement attempt and
interpose on subsequent requests to interfere with the reports. Although such
interference is plausible, censors do not generally interfere with measurements
today, so we leave this consideration to future work.

Attackers may attempt to submit poisoned measurement results to alter the
conclusions that \name{} draws about censorship. We could try to employ
reputation systems to thwart such attacks, although it would be practically
impossible to completely prevent such poisoning from untrusted
clients~\cite{Hao2009:snare}.

\balance\section{Conclusion}\label{sec:conclusion}

Despite the importance of measuring the extent and nature of Internet
censorship, doing so is difficult because it requires deploying a large number
of geographically diverse vantage points, and recruiting volunteers for such a
deployment is a significant deployment hurdle.  This paper presents an alternate
approach: rather than requiring users to install custom measurement software, we
take advantage of the fact that users' Web browsers can perform certain types of
cross-origin requests, which we can harness to induce measurements of
reachability to arbitrary third-party domains.  Although only a limited amount
of information about the success of these requests leaks across domains, even a
small amount of leakage turns out to be enough to permit inferences about the
reachability of higher-level Web resources, including both entire domains and
specific URLs.

\name{} shifts the deployment burden from clients to webmasters. We have
designed \name{} so that deployment is simple (in many cases, webmasters
only add one line to the main Web page source). We also point out that
many webmasters are typically interested in monitoring the reachability
of their sites from various client geographies and networks in any case, so
deployment incentives are well-aligned.

%\xxx{interplay/relationship with tools that provide block lists, and
%  tools that provide more details about censorship practices.}

Although the types of measurements \name{} can perform may be more definitive
than tools that rely on informal user reports (\eg, Herdict), \name{} may
draw far fewer conclusions about the scope and methods of censorship
than tools that measure
censorship methods in detail (\eg, OONI, Centinel).  Ultimately, censorship
measurement is a complex, moving target, and no single measurement method or
tool can paint a complete picture.  What is
sorely missing from the existing set of measurement tools, however, is a way to
characterize censorship practices in broad strokes, based on a sizeable
and continuous set of client measurements.  By filling this important hole in our
understanding, \name{} can help bridge the gap between diverse yet inconclusive
user reports and detailed yet narrow or short-term fine-grained measurements.

% - Feasibility of using Encore to measure more sites with more users is still
%   unknown.
% - Community must develop set of ethical standards.
% - In either case, this work provides value:
%   - If there are ethical ways to deploy Encore, then we get valuable new
%     measurements on the state of Web censorship worldwide.
%   - If not, then our work highlights weaknesses in cross origin security
%     policy that should be fixed.

The prospect of using \name{} to collect measurements from unsuspecting users
has already stirred controversy within the networking community and prompted a
wider dialogue on ethics of network measurement~\cite{www-workshop-ns-ethics15}.
Forthcoming guidelines for ethical measurement will hopefully help determine
whether we can deploy \name{} more widely. Our work is beneficial regardless: If
wider deployment is appropriate, this paper has explained how \name{} could
yield valuable insight on Web censorship at a previously unattainable scale; if
ethical concerns make further deployment infeasible, our work is evidence that
attackers could use tools like \name{} to place users in harm's way and that
perhaps cross-origin security policy should be strengthened to prevent such
attacks.

\label{lastpage}
\end{sloppypar}

%\pagebreak

\appendix
\section{Example of a measurement task}\label{sec:appendix}

This is a complete example of JavaScript code that runs in a client's Web
browser to measure Web filtering using cross-origin embedding of a hidden image.
It uses jQuery~\cite{www-jquery}. The coordination server minifies and
obfuscates the source code before sending it to a client.

See \url{http://goo.gl/l8GU0R} for a simple demo of \name{}'s cross-origin
request mechanism.

\begin{small}
\begin{verbatim}
var M = Object();

// A measurement ID is a unique identifier
// linking all submissions of a measurement.
M.measurementId = ...  // a UUID.

// This function embeds an image from a remote
// origin, hides it, and sets up callbacks to
// detect success or failure to load the image.
M.measure = function() {
  var img = $('<img>');
  img.attr('src', '//target/image.png');
  img.style('display', 'none');
  img.on('load', M.sendSuccess);
  img.on('error', M.sendFailure);
  img.appendTo('html');
}

// This function submits a result using
// a cross-origin AJAX request. The server
// must allow such cross-origin submissions.
M.submitToCollector = function(state) {
  $.ajax({
    url: "//collector/submit" +
         "?cmh-id=" + this.measurementId +
         "&cmh-result=" + state,
  });
}
M.sendSuccess = function() {
  M.submitToCollector("success");
}
M.sendFailure = function() {
  M.submitToCollector("failure");
}

// Submit to the server as soon as the client
// loads the page, regardless of the
// measurement result. This indcates which
// clients attempted to run the measurement,
// even if they don't submit a final result.
M.submitToCollector("init");

// Run the measurement when the page loads.
$(M.measure);
\end{verbatim}
\end{small}
\vspace{8pt}

\section{SIGCOMM Signing Statement}

A version of this paper was published in SIGCOMM 2015~\cite{Burnett2015} and is
accompanied by the following statement from the SIGCOMM Program Committee:

\fp{}The SIGCOMM 2015 PC appreciated the technical contributions made in this paper,
but found the paper controversial because some of the experiments the authors
conducted raise ethical concerns. The controversy arose in large part because
the networking research community does not yet have widely accepted guidelines
or rules for the ethics of experiments that measure online censorship. In
accordance with the published submission guidelines for SIGCOMM 2015, had the
authors not engaged with their Institutional Review Boards (IRBs) or had their
IRBs determined that their research was unethical, the PC would have rejected
the paper without review. But the authors did engage with their IRBs, which did
not flag the research as unethical. The PC hopes that discussion of the ethical
concerns these experiments raise will advance the development of ethical
guidelines in this area. It is the PC's view that future guidelines should
include as a core principle that researchers should not engage in experiments
that subject users to an appreciable risk of substantial harm absent informed
consent. The PC endorses neither the use of the experimental techniques this
paper describes nor the experiments the authors conducted.

%\vspace{-0.1in}
%\section*{Acknowledgments}
% Comments for people we need to ack in the final version

%% Bibliography

\small
\setlength{\parskip}{-1pt}
\setlength{\itemsep}{-1pt}
% \footnotesize % SPACE
\balance\bibliography{ref,rfc}
\bibliographystyle{abbrv}
%\bibliographystyle{abbrvnat_noaddr} % SPACE
%\theendnotes % ENDNOTES
}{% !onlyAbstract
}

\end{document}